\documentclass[11pt,russian,a4paper,twoside]{article}  
\usepackage[cp1251]{inputenc}                            
\usepackage{amsmath,amsfonts,amssymb,amscd,euscript}
\usepackage[russian]{babel}

\usepackage{graphicx}
\usepackage{psfrag}
\makeatletter

\def\titlerus{\thispagestyle{empty} { } \vspace{-5mm} \noindent
\raisebox{-37pt}[\headheight][0pt]{\vbox{ \hbox to \textwidth{\hfil
\scriptsize  ВЕСТНИК \; УДМУРТСКОГО \; УНИВЕРСИТЕТА\hfil }
\vspace{2pt} \hrule \vspace{8pt} \hbox to \textwidth{\series \hfil  \issue}
\vspace{30pt} \hbox{УДК \UDK} }} \vspace{ 30pt plus 6pt }}

\def\titleeng{\vspace{3ex} \hfill Поступила  в редакцию \datereceive \par \vspace{5ex} \par
\noindent \parbox{166mm
}{\small {\textbf {\textit {\autorseng}}} \par {\bf \articleseng} \par
\vspace{10pt} \par \annotationeng \par \vspace{7pt} \par {\it Keywords}: \keywordseng
\par \vspace{7pt} \par \noindent \small {Mathematical Subject Classifications}: \MSC }
\par \vspace{30pt} \par \small \noindent \contactinformation}

\def\annotationandkeywordsrus{\noindent {\small \annotationrus \par } \vspace{8pt}
\noindent {\small {\it Ключевые слова}: \keywordsrus} \par \vspace{10pt}}

\@addtoreset{equation}{section}
\renewcommand{\section}{\@startsection{section}{1}{0pt}{1.3ex
plus 1ex minus .1ex}{1.3ex plus .1ex}{\bf\,\S\,}}

\renewcommand{\@begintheorem}[2]{\begin{trivlist}
\item[\hspace{\labelsep}{\bf \mbox{~~~}#1\ #2.}]}
\renewcommand{\@opargbegintheorem}[3]{\begin{trivlist}
\item[\hspace{\labelsep}{\bf \mbox{~~~}#1\ #2 {\rm (#3).}}]}
\renewcommand{\@endtheorem}{\end{trivlist}}

\newtheorem{teo}{Теорема}

\newtheorem{df}{Определение}

\newcommand{\doc}{\mbox{Д о к а з а т е л ь с т в о}}
\renewcommand{\@evenfoot}{}
\renewcommand{\@oddfoot}{}

\newcommand{\re}{\vspace{-0.5em}}
\renewcommand*{\@biblabel}[1]{#1.\hfill}

\newcommand*{\CSep}{.\ }
\renewcommand{\@makecaption}[2]{%
  \vskip\abovecaptionskip
  \sbox\@tempboxa{{\bf #1\CSep}{#2}}%
  \ifdim \wd\@tempboxa >\hsize
  \begin{center}%
    {\footnotesize{\bf #1\CSep}{#2\par}}%
  \end{center}%
  \else
    \global \@minipagefalse
    \hb@xt@\hsize{\hfil\box\@tempboxa\hfil}%
  \fi
  \vskip\belowcaptionskip%
}

\renewcommand{\@evenhead}{\raisebox{0pt}[\headheight][0pt]{\vbox{\hbox to\textwidth{\thepage \strut \hfil
\text{\autorsrus} \hfil } \hrule \vspace{8pt} \hbox to \textwidth{\series \hfil  \issue}}}}

\renewcommand{\@oddhead}{\raisebox{0pt}[\headheight][0pt]{\vbox{\hbox to\textwidth{  \strut \hfil
\text{\articleshortname} \hfil \thepage} \hrule \vspace{8pt} \hbox to \textwidth{\series \hfil  \issue}}}}

\newcommand{\series}{МАТЕМАТИКА}

\newcommand{\issue}{2010. Вып.\,2} 

\newcommand{\autorsrus}{П.\,Е.~Рябов, М.\,П.~Харламов} 
\newcommand{\autorseng}{P.\,E.~Ryabov, M.\,P.~Kharlamov} 

\newcommand{\articleshortname}{Аналитическая классификация особенностей обобщенного случая Ковалевской}
\newcommand{\articleseng}{Analytic classification of singularities in the generalized Kowalevski case}

\newcommand{\UDK}{517.938.5+531.38} 
\newcommand{\MSC}{70E17, 70G40} 

\newcommand{\annotationrus}{В задаче о движении волчка Ковалевской в двойном поле
(случай интегрируемости
А.\,Г.~Реймана--М.\,А.~Семенова-Тян-Шанского) вычислен тип всех
критических точек отображения момента. }

\newcommand{\annotationeng}{%
In the problem of motion of the Kowalevski top on two constant
fields (the A.\,G.~Reyman--M.\,A.~Semenov-Tian-Shansky case) the
type of all critical points of the momentum map is calculated.}

\newcommand{\keywordsrus}{интегрируемые гамильтоновы системы, отображение момента,
бифуркационная диаграмма, тип невырожденной особенности.}
\newcommand{\keywordseng}{integrable Hamiltonian system, momentum map,
bifurcation diagram, type of non-degenerate singularity.}

\newcommand{\datereceive}{23.06.10} 

\newcommand{\contactinformation}{Рябов Павел Евгеньевич,
к.\,ф.-м.\,н., доцент, кафедра теории вероятностей и математической
статистики, Финансовый университет при Правительстве Российской
Федерации, 125993, Россия, г. Москва,  Ленинградский просп., 49, E-mail: orelryabov@mail.ru \\
Харламов Михаил Павлович, д.\,ф.-м.\,н., профессор, кафедра
математического моделирования, Волгоградская академия госслужбы,
400131, Россия, г. Волгоград, ул. Гагарина,\,8, E-mail:
mharlamov@vags.ru}

\def\bR{\mathbb{R}}

\newcommand {\ri} {\mathrm{i}}
\newcommand {\mstrut}{\vphantom{\bigl(}}
\newcommand {\mF}{\mathcal{F}}
\newcommand {\mP}{\mathcal{P}}
\newcommand {\ds}{\displaystyle}
\newcommand {\mA}{\mathcal{A}}
\newcommand {\mK}{\mathcal{K}}
\newcommand {\pov}{\Gamma}

\usepackage{array}
\begin{document}

\titlerus

\begin{flushleft}
{\bf \copyright { \textit { \autorsrus}} \\[2ex]
{АНАЛИТИЧЕСКАЯ КЛАССИФИКАЦИЯ ОСОБЕННОСТЕЙ\\ОБОБЩЕННОГО ВОЛЧКА КОВАЛЕВСКОЙ}
\footnote{Работа выполнена при финансовой поддержке РФФИ (грант
10--01--00043).} }
\end{flushleft}

\annotationandkeywordsrus

\begin{flushleft}{\bf{Введение}}\end{flushleft}
Задача о движении волчка Ковалевской в двойном поле сил описывается
уравнениями \cite{Bogo}
\begin{gather}\label{en1_1}
\begin{array}{lll}
\displaystyle{2 \, \dot \omega_1=\omega_2\omega_3+\beta_3,}
&\displaystyle{ \dot \alpha_1=\alpha_2\omega_3-\alpha_3\omega_2,} &
\displaystyle{\dot\beta_1=\beta_2\omega_3-\omega_2\beta_3,}\\
\displaystyle{2 \, \dot \omega_2 = - \omega_1 \omega_3-\alpha_3,} &
\displaystyle{\dot \alpha_2=\omega_1\alpha_3-\omega_3\alpha_1,}&
\displaystyle{\dot\beta_2=\omega_1\beta_3-\omega_3\beta_1,}\\
\displaystyle{\dot\omega_3=\alpha_2-\beta_1,}& \displaystyle{\dot
\alpha_3=\alpha_1\omega_2-\alpha_2\omega_1,} & \displaystyle{\dot
\beta_3=\beta_1\omega_2-\beta_2\omega_1.}
\end{array}
\end{gather}
Здесь $\boldsymbol\omega$\,--- вектор угловой скорости, $\boldsymbol
\alpha, \boldsymbol\beta$\,--- напряженности силовых полей. В
работе~\cite{Kh34} показано, что без ограничения общности силовые
поля можно считать взаимно ортогональными. Тогда геометрические
интегралы системы (\ref{en1_1}) запишутся в виде ($a \geqslant b
\geqslant 0$)
\begin{gather*}
|{\boldsymbol\alpha}|^2=a^2,\quad |{\boldsymbol\beta}|^2=b^2,\quad
{\boldsymbol\alpha}\cdot {\boldsymbol\beta}=0.
\end{gather*}
На заданном этими уравнениями многообразии
$\mP^6\subset{\bR}^9({\boldsymbol\omega},{\boldsymbol\alpha},{\boldsymbol\beta})$
скобки Пуассона, индуцированные скобками Ли--Пуассона на коалгебре
$L_9^*$ \cite{Bogo}, невырождены, и система (\ref{en1_1})
гамильтонова с тремя степенями свободы с гамильтонианом
\begin{gather*}
H=\frac{1}{2}(2\omega_1^2+2\omega_2^2+\omega_3^2)-\alpha_1-\beta_2.
\end{gather*}
При $b=0$ система (\ref{en1_1}) описывает движение твердого тела в
поле силы тяжести при условиях С.\,В.~Ковалевской на распределение
масс, а при $a=b$\,--- cлучай Х.\,М.~Яхья \cite{Yeh}. Эти предельные
значения обладают группой симметрий и сводятся к семейству
интегрируемых систем с двумя степенями свободы. Далее
предполагается, что $a>b>0$ (неприводимый случай). Для сокращения
записи используются также обозначения $p^2=a^2+b^2,$ $r^2=a^2-b^2.$

Функции
\begin{gather*}
\begin{array}{l}
K=(\omega_1^2-\omega_2^2+\alpha_1-\beta_2)^2+(2\omega_1\omega_2+\alpha_2+\beta_1)^2,\\
G=\left[\omega_1\alpha_1+\omega_2\alpha_2+\frac{1}{2} \alpha_3
\omega_3\right]^2 +
\left[ \omega_1 \beta_1 + \omega_2 \beta_2 + \frac{1}{2} \beta_3 \omega_3 \right]^2+\\
\phantom{G=}+\omega_3\left[(\alpha_2\beta_3-\alpha_3\beta_2)\omega_1+(\alpha_3\beta_1-\alpha_1\beta_3)\omega_2+
\frac{1}{2}(\alpha_1\beta_2-\alpha_2\beta_1)\omega_3\right]-\\
\phantom{K=}-\alpha_1b^2-\beta_2 a^2
\end{array}
\end{gather*}
вместе с $H$ образуют на $\mP^6$ полный инволютивный набор
интегралов системы (\ref{en1_1}). Интегралы $K$ и $G$ указаны в
\cite{Bogo} и \cite{BobReySem}.

Определим интегральное отображение ${\mF}: \mP^{6} \to {\bR}^3,$
полагая ${\mF}(x)=\bigl(G(x), K(x), H(x)\bigr).$ Отображение ${\mF}$
принято называть {\it отображением момента}.

Обозначим через $\mK$ совокупность всех критических точек
отображения момента, то есть точек, в которых $\operatorname{rk}
d{\mF}(x)<3.$ Множество критических значений $\Sigma={\mF }(\mK)
\subset{\bR}^3$ называется {\it бифуркационной диаграммой}.
Множество $\mK$ можно стратифицировать рангом отображения момента,
представив в виде объединения $\mK =\mK^0 \cup \mK^1 \cup \mK^2.$
Здесь $ \mK^q= {\{x \in \mP^6 | \operatorname{rk} d{\mF}(x)=q \},}$
а точки из $ \mK^q$ называются критическими {\it ранга} $q.$ В
соответствии с этим и диаграмма $\Sigma$ становится клеточным
комплексом $\Sigma =\Sigma^0 \cup \Sigma^1 \cup \Sigma^2.$ С другой
стороны, на практике бифуркационные диаграммы описываются в терминах
некоторых поверхностей в пространстве констант первых интегралов.
Уравнения этих поверхностей (неявные или параметрические) зачастую
можно получить даже не вычисляя самих критических точек, как
дискриминантные множества некоторых многочленов (например, исходя из
особенностей алгебраических кривых, ассоциированных с
представлениями Лакса). Такие поверхности будем обозначим через
$\pov _i.$ Тогда критическое множество $\mK$ оказывается
объединением естественным образом возникающих инвариантных множеств
$\mathcal{M}_i=\mK \cap \mF^{-1}(\pov_i).$ Если поверхность $\pov_i$
записана регулярным уравнением $\phi_i(g,k,h) = 0,$ то
$\mathcal{M}_i$ определится как множество критических точек
интеграла $\phi_i(G,K,H),$ лежащих на его нулевом уровне, а
вычисленные в точке $\mathcal{M}_i$ компоненты градиента функции
$\phi_i$ в подстановке значений интегралов $G,K,H$ дадут
коэффициенты равной нулю линейной комбинации дифференциалов
$dG,dK,dH.$ В точке трансверсального пересечения двух поверхностей
$\pov_i$ и $\pov_j$ получим две независимые равные нулю комбинации,
поэтому в точках соответствующего пересечения $\mathcal{M}_i \cap
\mathcal{M}_j$ ранг $\mF$ равен~1. Очевидно, что точки
трансверсального пересечения трех поверхностей (углы бифуркационной
диаграммы) оказываются порожденными точками с условием
$\operatorname{rk} \mF =0.$ Множества $\mathcal{M}_i$ с
индуцированной на них динамикой далее называем {\it критическими
подсистемами}.

Критические подсистемы и уравнения поверхностей $\pov_i$ в
рассматриваемой задаче найдены в работе \cite{Kh34}. Подробное
описание стратификации критического множества по рангу отображения
момента изложено в \cite{Kh36}. Там же в виде явных неравенств
указаны области существования движений на поверхностях $\pov_i$ ---
множества $\Sigma_i,$ составляющие бифуркационную диаграмму. Как
следствие построен атлас всех сечений диаграммы $\Sigma$ плоскостями
постоянной энергии, то есть найдены все бифуркационные диаграммы
$\Sigma(h)$ отображения $G{\times}K,$ ограниченного на
изоэнергетические поверхности $\{H=h\} \subset \mP^6.$ Этих
результаты в необходимом здесь объеме приведены ниже.

Критических подсистемы оказываются интегрируемыми гамильтоновыми
(почти всюду) системами с числом степеней свободы меньшим трех. Для
них, в свою очередь, определено индуцированное отображение момента.
Описание критических множеств и бифуркаций {\it внутри} критических
подсистем получено в работах \cite{Zot}, \cite{KhSav},
\cite{Kh2009}. Однако классификация точек множества $\mK$ по
отношению ко всей системе с тремя степенями свободы на $\mP^6$ не
проводилась. В данной работе исследуется тип невырожденных
особенностей полного отображения момента.

Напомним понятие невырожденной особенности \cite{BolFom}
применительно к системе с тремя степенями свободы. Пусть $x$\,---
критическая точка ранга $q \leqslant 2$ отображения момента на
$\mP^{6}.$ Если она является критической точкой некоторого интеграла
$\varphi,$ то линеаризация $A_\varphi$ векторного поля
$\operatorname{sgrad} \varphi$ в точке $x$ является элементом
отождествляемой с $\operatorname{sp}(6,{\bR})$ алгебры всех
симплектических операторов в касательном пространстве $T_x\mP^{6}.$
Заменой координат в образе отображения момента добьемся того, что
точка $x$ будет критической для первых $3-q$ компонент $g_j$
отображения момента и регулярной для оставшихся $q.$ Рассмотрим в
$T_x\mP^{6}$ подпространство $W,$ натянутое на $\operatorname{sgrad}
g_1,$\ldots, $\operatorname{sgrad} g_{3-q},$ и его косоортогональное
дополнение $W^\prime.$ Тогда ${W \subset \operatorname{Ker}_{g_j}}$
и ${\operatorname{Im} A_{g_j} \subset W^\prime}.$ На
факторпространстве $W^\prime/W$ индуцируется симплектическая
структура, а операторы $A_{g_j}|_{W^\prime/W}$ являются элементами
алгебры Ли ${\operatorname{sp}(2(3-q),{\bR})}.$ Обозначим через
$\mA(x,{\mF})$ порожденную ими коммутативную подалгебру.

\begin{df}
Критическая точка $x\in \mK^q$ отображения момента $\mF $ называется
{\it невырожденной ранга} $q$ ({\it коранга} $3-q),$ если
$\mA(x,{\mF})$\,--- подалгебра Картана в
$\operatorname{sp}(2(3-q),{\bR}).$
\end{df}

На практике для проверки картановости необходимо найти указанную
замену в образе $\mF$ и убедиться в выполнении следующих условий:
операторы $A_{g_j}$ $(j=1,\ldots,3-q)$ линейно независимы, и в их
линейной оболочке найдется оператор $A,$ у которого $2(3-q)$
собственных значения различны (индуцированный им оператор в
$\mA(x,{\mF})$ называется {\it регулярным элементом}).

Известно, что собственные числа симплектического оператора разбиты
на группы трех типов --- пары вида $\pm \ri \mu$ (центр,
эллиптический тип), $\pm \mu$ (седло, гиперболический тип) и
четверки $\pm \mu_1 \pm \ri \mu_2$ (фокусная особенность), где $\mu,
\mu_1, \mu_2 \in \bR$ и для регулярного элемента все они отличны от
нуля. Здесь и далее $\ri$\,--- мнимая единица. Количество элементов
в каждой такой группе обозначим соответственно через $m_1, m_2,
m_3$. Для невырожденной критической точки ранга $q$ имеем
$m_1+m_2+2m_3=3-q.$ Четверка $(q,m_1,m_2,m_3)$ называется типом
невырожденной критической точки. После того как тип критической
точки определен, слоение Лиувилля в ее окрестности для
вещественно-аналитических систем оказывается локально
симплектоморфно некоторому модельному слоению (подробности см. в
\cite{BolFom}).

Перейдем к описанию критических точек обобщенного волчка
Ковалевской.

\begin{flushleft}
\noindent{\bf{\S\,1.\,Критические точки ранга 0}} \end{flushleft}
Особенность нулевого ранга предполагает, в частности, равенство
$dH=0,$ что возможно лишь в неподвижных точках гамильтоновой
системы. Здесь их ровно четыре:
\begin{gather}\label{en2_1}
c_k: \; {\boldsymbol\omega}={\boldsymbol 0}, \quad  {\boldsymbol
\alpha} = (\varepsilon_1  a ,\, 0,\, 0), \quad {\boldsymbol \beta}=
(0,\, \varepsilon_2 b, \, 0),
\end{gather}
где $\varepsilon_1^2=\varepsilon_2^2=1$ и, соответственно,
$k=1,\ldots,4.$ Значения первых интегралов образуют остов $\Sigma^0$
бифуркационной диаграммы --- четыре точки в $\bR^3(h,k,g)$:
\begin{gather*}
P_k:\quad g=-ab(\varepsilon_2 a+\varepsilon_1 b), \quad
k=(\varepsilon_1 a-\varepsilon_2 b)^2, \quad h=-\varepsilon_1
a-\varepsilon_2 b.
\end{gather*}
Упорядочим $c_k$ по возрастанию $h.$

В работе \cite{KhZot} найден индекс Морса гамильтониана $H$ в
неподвижных точках, что в значительной мере определяет характер
поведения системы в их окрестности. Однако строгая классификация
требует указания типа этих точек как критических точек отображения
момента.
\begin{teo}
{\it Точки $c_k$ $(k=1,\ldots,4)$ являются невырожденными
особенностями ранга~$0$ отображения момента ${\mF}.$ При этом $c_1$
имеет тип $(0,3,0,0)$ {\rm (}центр-центр-центр{\rm )}, $c_2$\,---
тип $(0,2,1,0)$ {\rm (}центр-центр-седло{\rm )}, $c_3$\,--- тип
$(0,1,2,0)$ {\rm (}центр-седло-седло{\rm )}, $c_4$\,--- тип
$(0,0,3,0)$ {\rm (}седло-седло-седло{\rm )}.}
\end{teo}

\doc.
Независимость операторов $A_H, A_K, A_G,$ полученных линеаризацией
полей $\operatorname{sgrad} H,$ $\operatorname{sgrad} K $ и
$\operatorname{sgrad} G$ в точках (\ref{en2_1}), легко проверяется
непосредственно в пространстве $\bR^9.$ В этом же пространстве
характеристический многочлен оператора $A_H$ с точностью до
постоянного множителя имеет вид:
\begin{gather*}
\mu^3 [\mu^2 + \varepsilon_1  a+\varepsilon_2  b][2\mu^2 +
\varepsilon_1  a][2\mu^2 + \varepsilon_2  b].
\end{gather*}
Трехкратный нулевой корень отвечает геометрическим интегралам, а
оставшийся многочлен шестой степени, очевидно, не имеет кратных
корней. Поэтому $A_H$ является регулярным элементом соответствующей
подалгебры. Утверждение теоремы следует из предположения о
неприводимости. \hfill $\square$

\begin{flushleft}{\bf{\S\,2.\,Невырожденные особенности ранга
1}}\end{flushleft} В составе множества $\mK^1$ имеются следующие
шесть семейств маятниковых движений \cite{Kh34} (первый индекс
соответствует верхнему знаку):
\begin{gather*}
\begin{array}{l}
{\mathcal{L}}_{1,2}=\{{\boldsymbol \alpha } \equiv \pm a{\bf e}_1,
\; {\boldsymbol \beta } = b({\bf e}_2 \cos \theta - {\bf e}_3 \sin
\theta ), \; {\boldsymbol \omega } = \theta ^ {\boldsymbol \cdot}
{\bf e}_1 , \; 2\theta ^{ {\boldsymbol \cdot}  {\boldsymbol \cdot} }
=  - b\sin \theta\},\\
{\mathcal{L}}_{3,4}=\{{\boldsymbol \alpha } = a({\bf e}_1 \cos
\theta + {\bf e}_3 \sin \theta ), \; {\boldsymbol \beta } \equiv
\pm b{\bf e}_2 , \; {\boldsymbol \omega } = \theta ^ {\boldsymbol
\cdot}  {\bf e}_2 , \; 2\theta ^{ {\boldsymbol \cdot}  {\boldsymbol
\cdot} }  = -
a\sin \theta\}, \\
{\mathcal{L}}_{5,6}=\{{\boldsymbol{\alpha }} = a({\bf{e}}_1 \cos
\theta - {\bf{e}}_2 \sin \theta ),\;{\boldsymbol{\beta }} =  \pm
b({\bf{e}}_1 \sin \theta  + {\bf{e}}_2 \cos \theta ), \;
{\boldsymbol{\omega }} = \theta ^ {\boldsymbol \cdot}  {\bf{e}}_3
,\; \theta ^{ {\boldsymbol \cdot} {\boldsymbol \cdot} }  =  - (a \pm
b)\sin \theta\}.
\end{array}
\end{gather*}
Из них, естественно, следует исключить неподвижные точки
(\ref{en2_1}), принадлежащие всем шести семействам и отвечающие
четырем значениям энергии $h \in \mathcal{H}=\{\pm(a\pm b)\}.$
Семействам ${\mathcal{L}}_j$ отвечают значения первых интегралов,
заполняющие кривые $\lambda_j$ в составе одномерного остова
бифуркационной диаграммы (из них также исключены образы неподвижных
точек):
\begin{gather*}
\begin{array}{lll}
\lambda_{1,2}=\{g = a^2 h\pm a r^2,& k=(h\pm
2a)^2, & h > \mp(a\pm b), \; h \notin \mathcal{H} \},\\
\lambda_{3,4}=\{g = b^2 h\mp b r^2, & k=(h\pm
2b)^2,& h > -(a\pm b),\; h \notin \mathcal{H}\},\\
\lambda_{5,6}=\{g=\pm abh,& k=(a\mp b)^2,& h > -(a\pm b),\; h \notin
\mathcal{H}\}.
\end{array}
\end{gather*}
Оставшуюся часть множества $\mK^1$ составляют критические движения
случая Богоявленского \cite{Bogo}. Они аналитически организованы в
три семейства периодических решений, ниже обозначаемых через
${\mathcal{L}}_7-{\mathcal{L}}_9$ (топологически эти семейства
заполняют в $\mP^6$ четыре связных двумерных многообразия). Первое
описание этих траекторий дано в \cite{Zot}. Здесь удобно
воспользоваться параметризацией этих семейств, указанной в
\cite{Kh37}. Она представляет собой алгебраические выражения
исходных фазовых переменных через одну вспомогательную переменную
$\theta,$ зависимость которой от времени выражается стандартными
эллиптическими функциями, при том, что и постоянные всех интегралов
также явно выражены через одну постоянную $s.$ Имеем следующие
выражения для фазовых переменных:
\begin{gather*}
\begin{array}{l}
\omega_1=\ds{-\frac{1}{r} \sqrt{\frac{r_1 r_2 \psi_1}{2 s}}},\quad
\omega_2=-\ds{\frac{1}{r} \sqrt{-\frac{r_1 r_2 \psi_2}{2 s}}},\quad
\omega_3= \ds{\frac{\theta}{r_1+r_2} \sqrt{\frac{2 r_1
r_2}{s}}},\\[4mm]
\alpha_1=-s -\ds{\frac{r_1 r_2 \bigl[r^2+(r_1+r_2)^2
\bigr](\psi_1+\psi_2)}{4r^2(r_1+r_2)^2 s}},\quad
\alpha_2=-\ds{\frac{r_1 r_2 \bigl[r^2+(r_1+r_2)^2
\bigr]\sqrt{-\psi_1 \psi_2}}{2 r^2
(r_1+r_2)^2 s}},\\[4mm]
\beta_1=\ds{\frac{r_1 r_2 \bigl[r^2-(r_1+r_2)^2\bigr]\sqrt{-\psi_1
\psi_2}}{2 r^2 (r_1+r_2)^2 s}},\quad \beta_2=-s-\ds{\frac{r_1 r_2
\bigl[r^2-(r_1+r_2)^2\bigr](\psi_1+\psi_2)}{4 r^2
(r_1+r_2)^2 s}},\\[4mm]
\alpha_3=\ds{\frac{r_1 \sqrt{-\psi_1}}{r}},\quad
\beta_3=\ds{\frac{r_2\sqrt{-\psi_2}}{r}}.
\end{array}
\end{gather*}
Здесь $\psi_j  = \theta^2 - 2s \ds{\frac{r_1+r_2}{r_j}}\theta  -
(r_1+r_2)^2$ $(j=1,2)$ и для различных семейств следует положить
\begin{gather*}
\begin{array}{ll}
{\mathcal{L}}_7: s \in [ - b,0), & r_1  = \sqrt {\mstrut a^2 - s^2 }
> 0, \quad r_2 = \sqrt {\mstrut b^2  - s^2 } \geqslant 0, \\[3mm]
{\mathcal{L}}_8: s \in (0,b], & r_1  = \sqrt {\mstrut a^2  - s^2 }
> 0,
\quad r_2  =  - \sqrt {\mstrut b^2  - s^2 }  \leqslant 0, \\[3mm]
{\mathcal{L}}_9: s \in [a, + \infty ), & \left\{\begin{array}{l} r_1
= \ri \, r_1^*, \quad r_2  = \ri \,r_2^* \\ 0 \leqslant r_1^*  =
\sqrt {\mstrut s^2  - a^2 } < r_2^*  = \sqrt {\mstrut s^2 - b^2 }
\end{array}.\right.
\end{array}
\end{gather*}
Значения первых интегралов на этих движениях определяют еще одну
часть одномерного остова бифуркационной диаграммы $\Sigma$ в виде
трех кривых $\delta_i \subset \bR^3$ $(i=1,2,3),$ параметрические
уравнения которых получены в \cite{Kh36}
\begin{gather*}
\begin{array}{l}
\delta_1: \left\{
\begin{array}{l}
\displaystyle{h=2s-\frac{1}{s}\sqrt{(a^2-s^2)(b^2-s^2)}   }\\[2mm]
\displaystyle{f^2={-\frac{2}{s}\sqrt{(a^2-s^2)(b^2-s^2)}}(\sqrt{a^2-s^2}+
\sqrt{b^2-s^2})^2}
\end{array} \right. , \quad s \in [-b,0); \\
\delta_2: \left\{
\begin{array}{l}
\displaystyle{h=2s+\frac{1}{s}\sqrt{(a^2-s^2)(b^2-s^2)}   }\\[2mm]
\displaystyle{f^2={\frac{2}{s}\sqrt{(a^2-s^2)(b^2-s^2)}}(\sqrt{a^2-s^2}-
\sqrt{b^2-s^2})^2}
\end{array} \right. , \quad s \in (0,b]; \\
\delta_3: \left\{
\begin{array}{l}
\displaystyle{h=2s-\frac{1}{s}\sqrt{(s^2-a^2)(s^2-b^2)}   }\\[2mm]
\displaystyle{f^2={\frac{2}{s}\sqrt{(s^2-a^2)(s^2-b^2)}}(\sqrt{s^2-b^2}-
\sqrt{s^2-a^2})^2}
\end{array} \right., \quad  s \in [a,+\infty).
\end{array}
\end{gather*}
Здесь $k \equiv 0,$ $f=\sqrt{\mstrut 2(p^2h-2g)}$\,--- значение
частного интеграла О.\,И.~Богоявленского \cite{Bogo}
\begin{gather}\label{en3_1}
F=(\omega_1^2+\omega_2^2)\omega_3+2(\alpha_3 \omega_1+\beta_3
\omega_2).
\end{gather}

Пусть $h_0$\,--- минимальное значение энергии на кривой $\delta_3.$
Оно отвечает единственному на полупрямой $s>a$ корню $s_0$ уравнения
\begin{gather}\label{en3_2}
s^4-a^2 b^2-2 s^2\sqrt{\mathstrut s^2-a^2}\sqrt{\mathstrut
s^2-b^2}=0.
\end{gather}

\begin{teo}
{\it Критические точки ранга 1 невырождены, за исключением следующих
значений энергии: на $\mathcal{L}_2$ $h = 2a ,
\ds{\frac{3a^2+b^2}{2a}}$; на $\mathcal{L}_3$ $h = -2b $; на
$\mathcal{L}_4$ $h = 2b , \ds{\frac{a^2+3b^2}{2b}}$; на
$\mathcal{L}_5$ $h =\pm 2\sqrt{a b}$; на $\mathcal{L}_9$ $h = h_0.$
В зависимости от значений параметров $h,s$ тип невырожденных
критических точек в $\mP^6$ определяется из табл.~1.}
\end{teo}
{\renewcommand{\arraystretch}{1.5} \setlength{\extrarowheight}{-2pt}
\begin{table}[ht]
\centering
\begin{tabular}{|c|l|l|}
\multicolumn{3}{r}{\small{Таблица 1}}\\
\hline $\mK^1$& \multicolumn{1}{c|}{Образ в
 ${\bR}^3(h,k,g)$}&\multicolumn{1}{c|}{Тип в
 $\mP^6$}\\
\hline ${\mathcal{L}}_1$ &\begin{tabular}{l}
$\lambda_1=\lambda_{11}\cup\lambda_{12},$\\
$\lambda_{11}:-(a+b)<h<-(a-b),$\\
$\lambda_{12}:h>-(a-b),$
\end{tabular}
&\begin{tabular}{l} $(1,2,0,0)$
(центр-центр)\end{tabular}\\
\hline ${\mathcal{L}}_2$ &\begin{tabular}{l}
$\lambda_2=\lambda_{21}\cup\lambda_{22}\cup\lambda_{23},$\\
$\lambda_{21}:a-b<h<a+b,$\\
$\lambda_{22}:a+b<h<2a,$\, $\ds{h\ne\frac{3a^2+b^2}{2a}},$\\
$\lambda_{23}: h>2a$
\end{tabular}&\begin{tabular}{l}
$\lambda_{21},\lambda_{22}: (1,0,2,0)$
(седло-седло)\\
$\lambda_{23}: (1,1,1,0)$ (центр-седло)
\end{tabular} \\
\hline ${\mathcal{L}}_3$ &
\begin{tabular}{l}
$\lambda_3=\lambda_{31}\cup\lambda_{32}\cup\lambda_{33},$\\
$\lambda_{31}:-(a+b)<h<-2b,$\\
$\lambda_{32}:-2b<h<a-b,$\\
$\lambda_{33}: h>a-b$
\end{tabular}&\begin{tabular}{l}
$\lambda_{31}: (1,2,0,0)$ (центр-центр)\\
$\lambda_{32},\lambda_{33}: (1,1,1,0)$ (центр-седло)
\end{tabular}
\\
\hline ${\mathcal{L}}_4$ &
\begin{tabular}{l}
$\lambda_4=\lambda_{41}\cup\lambda_{42}\cup\lambda_{43},$\\
$\lambda_{41}:-(a-b)<h<2b,$\\
$\lambda_{42}:2b<h<a+b,$\\
$\lambda_{43}: h>a+b,$\, $\ds{h\ne\frac{a^2+3b^2}{2b}}$
\end{tabular}&\begin{tabular}{l}
$\lambda_{41}:(1,1,1,0)$ (центр-седло)\\
$\lambda_{42},\lambda_{43}: (1,0,2,0)$ (седло-седло)
\end{tabular}\\
\hline ${\mathcal{L}}_5$ &
\begin{tabular}{l}
$\lambda_5=\lambda_{50}\cup\left(\bigcup_{k=1}^4\,\lambda_{5k}\right),$\\
$\lambda_{51}:-(a+b)<h<-2\sqrt{ab}$\\
$\lambda_{50},-2\sqrt{ab}<h<2\sqrt{ab}$\\
$\lambda_{52}, \lambda_{53},\lambda_{54}:h>2\sqrt{ab}$
\end{tabular}&\begin{tabular}{l}
$\lambda_{51}:(1,2,0,0)$ (центр-центр)\\
$\lambda_{50}:(1,0,0,1)$ (фокус-фокус)\\
$\lambda_{52}, \lambda_{53},\lambda_{54}: (1,0,2,0)$ (седло-седло)
\end{tabular}
\\
\hline  ${\mathcal{L}}_6$ & \begin{tabular}{l}$\lambda_6:h>-(a-b)$
\end{tabular}&\begin{tabular}{l} $(1,1,1,0)$
(центр-седло)\end{tabular}\\
\hline ${\mathcal{L}}_7$ & \begin{tabular}{l}$\delta_1:s\in(-b,0)$
\end{tabular}&\begin{tabular}{l} $(1,2,0,0)$
(центр-центр)\end{tabular}\\
\hline ${\mathcal{L}}_8$ & \begin{tabular}{l}$\delta_2:s\in(0,b)$
\end{tabular}&\begin{tabular}{l} $(1,1,1,0)$
(центр-седло)\end{tabular}\\
\hline ${\mathcal{L}}_9$ &\begin{tabular}{l}
$\delta_3=\delta_{31}\cup\delta_{32}$\\
$\delta_{31}: s\in(a,s_0)$\\
$\delta_{32}: s\in(s_0,+\infty)$
\end{tabular}
&\begin{tabular}{l} $\delta_{31}:(1,1,1,0)$
(центр-седло)\\
$\delta_{32}: (1,2,0,0)$ (центр-центр)
\end{tabular}\\
\hline
\end{tabular}\,
\end{table}
}

\doc. Рассмотрим произвольную точку $x_0\in{\mathcal{L}}_k.$ Определим следующие
функции
\begin{gather*}
\begin{array}{lll}
({\mathcal{L}}_{1,2})& g_1=K-2(h\pm 2a)H, & g_2=G-a^2H,\\
({\mathcal{L}}_{3,4})& g_1=K-2(h\pm 2b)H, & g_2=G-b^2H,\\
({\mathcal{L}}_{5,6})& g_1=\pm abH-G, & g_2=K,\\
({\mathcal{L}}_{7,8,9})&g_1=2G-(p^2-\tau)H, & g_2=K.
\end{array}
\end{gather*}
Положим также $g_3=H.$ Поскольку все неподвижные точки имеют ранг 0
и уже исключены, то $dg_3(x_0) \ne 0.$ Выбранные функции $g_1,$
$g_2$ в точке $x_0$ имеют особенность: $dg_1(x_0)=dg_2(x_0)=0.$
Линеаризации векторных полей $\operatorname{sgrad} g_k$ $(k=1,2)$ в
точке $x_0$ дают линейные симплектические операторы $A_{g_k}:
T_{x_0}\mP^6 \to T_{x_0}\mP^6.$ Непосредственно проверяется, что они
линейно независимы, то есть порождают подалгебру в
$\operatorname{sp}(6,{\bR})$ размерности 2. Характеристическое
уравнение для оператора $A_{g_1}$ имеет два нулевых корня: $\ker
A_{g_1}=T_{x_0}\mathcal{L}_k.$ Остальная часть характеристического
многочлена имеет вид
\begin{gather*}
\begin{array}{ll}
{\mathcal{L}}_{1,2}:&[\mu^2+4(a^2-b^2)(h\pm 2a)]\,[\mu^2\pm 8a(h\pm
2a)^2]=0,\\
{\mathcal{L}}_{3,4}:&[\mu^2-4(a^2-b^2)(h\pm 2b)]\,[\mu^2\pm 8b(h\pm
2b)^2]=0,\\
{\mathcal{L}}_{5,6}: & 4\mu^4\mp 2abh(a\mp
b)^2\mu^2\pm b^3a^3(a\mp b)^4=0,\\
{\mathcal{L}}_{7,8,9}: &\mu^4+u_{k}\mu^2+v_k=0 \qquad (k=1,2,3),
\end{array}
\end{gather*}
где коэффициенты $u_k, v_k$ определяются по формулам
\begin{gather*}
\begin{array}{lcl}
u_{1,2}&=& - \ds{\frac{2}{s}}\bigl(\sqrt{\mathstrut {{a^2-s^2}}}\pm
\sqrt{\mathstrut {{b^2-s^2}}}\bigr)^2 \Bigl[a^2b^2-s^4 \pm
6s^2\sqrt{\mathstrut{{(a^2-s^2)(b^2-s^2)}}}\Bigr],\\
v_{1,2}&= &\pm
\ds{16\sqrt{\mathstrut{{(a^2-s^2)(b^2-s^2)}}}\bigl(\sqrt{\mathstrut
{{a^2-s^2}}}\pm \sqrt{\mathstrut {{b^2-s^2}}}\bigr)^4
}\Bigr[a^2b^2-s^4\pm 2s^2\sqrt{\mathstrut {{(a^2-s^2)(b^2-s^2)}}}\Bigl], \\
u_3&=& \phantom{-} \ds{\frac{2}{s}}\bigl(\sqrt{\mathstrut
s^2-a^2}-\sqrt{\mathstrut s^2-b^2}\bigr)^2 \Bigl[a^2b^2-s^4+
6s^2\sqrt{\mathstrut(s^2-a^2)(s^2-b^2)}\Bigr],\\
v_3&=&\phantom{\pm}\ds{16\sqrt{\mathstrut(s^2-a^2)(s^2-b^2)}\bigl(\sqrt{\mathstrut
s^2-a^2}-\sqrt{\mathstrut s^2-b^2}\bigr)^4
}\Bigr[a^2b^2-s^4+2s^2\sqrt{\mathstrut (s^2-a^2)(s^2-b^2)}\Bigl].
\end{array}
\end{gather*}

В табл.~1 кривые $\lambda_1 - \lambda_5, \delta_3$ разбиты на
участки значениями $h \in \mathcal{H},$ а также такими $h,$ при
которых нарушается условие отсутствия кратных корней у
соответствующего характеристического многочлена. На каждом из таких
участков все корни соответствующего характеристического уравнения
различны и разбиваются на пары, определяющие тип невырожденной
особенности. Для примера рассмотрим кривую $\delta_3$ $(s>a).$
Характеристическое уравнение в точках $\mathcal{L}_9$ относительно
$\mu^2$ имеет корни
\begin{gather*}
\begin{array}{l}
\mu^2_{(1)}= - 8s \sqrt{\mathstrut s^2-a^2}\sqrt{\mathstrut s^2-b^2}
\bigl(\sqrt{\mathstrut s^2-a^2}-\sqrt{\mathstrut s^2-b^2}\bigr)^2 <
0,\\
\mu^2_{(2)}=\ds{\frac{2}{s}}\bigl(\sqrt{\mathstrut
s^2-a^2}-\sqrt{\mathstrut s^2-b^2}\bigr)^2\bigl(s^4-a^2 b^2-2
s^2\sqrt{\mathstrut s^2-a^2}\sqrt{\mathstrut s^2-b^2}\bigr).
\end{array}
\end{gather*}
Согласно (\ref{en3_2}), последнее выражение меняет знак при переходе
через значение $s_0$: $\mu^2_{(2)}>0$ при $s\in (a,s_0)$ и
$\mu^2_{(2)}< 0$ при $s > s_0.$ В первом случае получаем особенность
центр-седло, во втором --- центр-центр. \hfill $\square$

Отметим, что согласно табл.~1 получено аналитическое доказательство
существования невырожденной особенности фокусного типа на части
множества $\mathcal{L}_5,$ которая отображается в $\lambda_{50}$
(значения энергии ${-2\sqrt{ab}<h<2\sqrt{ab}}$). В изоэнергетических
сечениях $\Sigma(h)$ имеем изолированную точку на бифуркационной
диаграмме.

\begin{flushleft}{\bf{\S\,3.\,Невырожденные особенности ранга 2}}\end{flushleft}
Как показано в \cite{Kh34}, множество $\mK^2$ всех критических точек
ранга $2$ есть объединение трех критических подсистем
$\mathcal{M}_i=\mK \cap \mF^{-1}(\pov_i)$ $(i=1,2,3)$ за вычетом уже
исследованных точек~${\mK^0 \cup \mK^1}$. Поверхности $\pov_i$
таковы:
\begin{gather}\label{en4_1}
\begin{array}{l}
\pov_1=\{k=0\};\\[4mm]
\pov_2=\{\ds{h = \frac {\ell^2-1}{2m}-p^2m}, \; \ds{k= r^4 m^2} , \;
\ds{g = \frac{(\ell^2-1)p^2}{4 m}-
\frac{p^4+r^4}{2}m}\};\\[4mm]
\pov_3=\{ \ds{h = \frac {p^2 -\tau}{2s}+ s}, \; \ds{k= \frac{\tau^2
- 2p^2\tau + r^4}{4s^2}+\tau}, \; \ds{g = \frac{p^4-r^4}{4s}+
\frac{1}{2}(p^2 -\tau)s}\}.
\end{array}
\end{gather}
Здесь величины $m,\ell,s,\tau$, служащие параметрами на
поверхностях, являются произвольными постоянными частных интегралов
соответствующих подсистем \cite{KhSav,Kh34}.

Первая подсистема $\mathcal{M}_1$ уже упоминалась выше как случай
частной интегрируемости, найденный О.\,И.~Богоявленским \cite{Bogo}.
Отметим сразу, что для этой системы каждый регулярный относительно
нее тор ${\mathbb{T}}^2\in \{x\in \mathcal{M}_1: H=h,F=f\},$ где
$F$\,--- интеграл (\ref{en3_1}), является \textit{эллиптическим} по
отношению к $\mP^6.$ Это вытекает из того, что интеграл $K$ есть
положительная всюду функция и обращается в ноль на $\mathcal{M}_1.$

\begin{teo} {\it Все критические точки ранга 2 на многообразии $\mathcal{M}_1$
являются невырожденными типа $(2,1,0,0)$ за исключением точек
нулевого уровня интеграла $F.$}
\end{teo}

\doc. Напомним, что на $\mathcal{M}_1$ нет критических точек интеграла $H$
\cite{Zot}, и отметим, что кроме множеств $\mathcal{L}_7 -
\mathcal{L}_9$ на $\mathcal{M}_1$ нет точек зависимости интегралов
$H$ и $G.$ Последнее вытекает из результатов \cite{Kh34}. Таким
образом, эти два интеграла регулярны и независимы на $\mathcal{M}_1
\cap \mK^2.$ В то же время всюду на $\mathcal{M}_1$ имеем $dK=0.$
Характеристическое уравнение оператора $A_K$ в $\bR^9$ при условии
$K=0$ легко выписывается, имеет семь нулевых корней, а оставшийся
сомножитель $\mu^2+4 f^2$ имеет два различных мнимых корня при $f\ne
0,$ что и доказывает теорему. \hfill $\square$

Образом множества $\{F=0\}$ при отображении момента в $\bR^3(h,k,g)$
является луч
\begin{gather*}
\Delta_1= \left\{k=0, \, 2g=p^2h, \, h \geqslant -2b\right\}.
\end{gather*}
В силу теоремы это множество соответствует наличию вырожденных
критических точек.

На многообразии $\mathcal{M}_2,$ найденном в работе \cite{Kh32},
система (\ref{en1_1}) имеет явное алгебраическое решение
\cite{KhSav}:
\begin{gather}\label{en4_2}
\begin{array}{l}
\displaystyle{\alpha _1  = \frac{\mathstrut 1} {{2(s_1  - s_2 )^2
}}[(s_1 s_2
- a^2 )\psi + S_1 S_2 \varphi _1 \varphi _2 ], }\\
\displaystyle{\alpha _2  = \frac{\mathstrut 1} {{2(s_1  - s_2 )^2
}}[(s_1 s_2
- a^2)\varphi _1 \varphi _2  - S_1 S_2 \psi ], }\\
\displaystyle{\beta _1  =  - \frac{\mathstrut 1} {{2(s_1  - s_2 )^2
}}[(s_1
s_2  - b^2)\varphi _1 \varphi _2  - S_1 S_2 \psi ], }\\
\displaystyle{\beta _2  = \frac{\mathstrut 1} {{2(s_1  - s_2 )^2
}}[(s_1 s_2
- b^2 )\psi + S_1 S_2 \varphi _1 \varphi _2 ], }\\
\displaystyle{\alpha _3  = \frac{\mathstrut r} {{s_1  - s_2 }}S_1
,\quad
\beta _3  =\frac{r} {{s_1  - s_2 }}S_2, }\\
\displaystyle{\omega _1  = \frac{\mathstrut r} {{2(s_1  - s_2
)}}(\ell - 2ms_1 )\varphi _2,\quad \omega _2  = \frac{r} {{2(s_1 -
s_2
)}}(\ell  - 2ms_2)\varphi _1, }\\
\displaystyle{\omega _3  = \frac{\mathstrut 1} {{s_1  - s_2 }}(S_2
\varphi _1 - S_1 \varphi _2 )}.
\end{array}
\end{gather}
Здесь обозначено
\begin{gather}\label{en4_3}
\begin{array}{l}
\displaystyle{\psi = 4ms_1 s_2  - 2\ell (s_1  + s_2 ) +
\frac{1}{m}(\ell ^2  - 1)},\quad  \displaystyle{\varphi (s) = 4ms^2
- 4\ell s + \frac{1} {m}(\ell ^2
- 1)},\\[4mm]
\displaystyle{S_1 = \sqrt {s_1^2 - a^2},}\quad
\displaystyle{\varphi_1 = \sqrt {\mstrut - \varphi (s_1)},}\quad
\displaystyle{S_2 = \sqrt {b^2 - s_2^2},}\quad
\displaystyle{\varphi_2 = \sqrt {\mstrut\varphi (s_2)}.}
\end{array}
\end{gather}

Особое множество $\Delta_1$ получает здесь уравнение $m=0.$
Соответствующая особенность в выражениях (\ref{en4_3}) устранима,
так как на $\pov_2$ имеется очевидное тождество
\begin{gather*}
2 p^2 m^2+ 2 h m+1=\ell^2.
\end{gather*}
Другая, топологическая, особенность отмечена в \cite{KhSav} и
связана с возникновением дополнительной симметрии при $\ell=0, m<0.$
В пространстве $\bR^3(h,k,g)$ имеем, в соответствии с (\ref{en4_1}),
особое множество
\begin{gather*}
\Delta_2= \left\{k=\ds{\frac{1}{r^4}(2g-p^2h)^2}, \,
g=\ds{\frac{1}{4p^2}} [(2p^4-r^4)h \pm r^4 \sqrt{\mathstrut
h^2-2p^2},\,h \geqslant p\sqrt{\mathstrut 2}\right\}.
\end{gather*}

\begin{teo} {\it За исключением точек, лежащих в прообразе кривых $\Delta_1,
\Delta_2,$ все критические точки ранга 2 на многообразии
$\mathcal{M}_2$ невырождены. Они имеют тип $(2,1,0,0)$ для $m>0$ и
$(2,0,1,0)$ для $m<0.$}
\end{teo}
\doc.
Исключая $\ell,m$ из уравнений $\pov_2,$ получим уравнение этой
поверхности в виде $(2g-p^2h)^2-r^4 k=0.$ Поэтому в качестве
единственного интеграла, имеющего особенность в каждой точке
$\mathcal{M}_2 \cap \mK^2,$ удобно взять функцию:
\begin{gather*}
\Phi = (2G-p^2H)^2-r^4 K.
\end{gather*}
Характеристическое уравнение оператора $A_\Phi$ после необходимой
факторизации по нулевому корневому подпространству в подстановке
явных выражений (\ref{en4_2}) примет вид
\begin{gather*}
\mu^2+4r^{12} \ell^2 m=0,
\end{gather*}
и за исключением случаев $m=0$ ($\Delta_1$) и $\ell=0$ ($\Delta_2$)
имеет два различных корня, чисто мнимых при $m>0$ и вещественных при
$m<0.$ Теорема доказана. \hfill $\square$

На многообразии $\mathcal{M}_3,$ найденном в работе \cite{Kh34},
явное алгебраическое решение системы (\ref{en1_1}) указано в
\cite{Kh2009}. Для заданных констант $s,\tau$ введем также
обозначения
\begin{gather*}
\begin{array}{l}
\sigma = \tau^2-2p^2 \tau+r^4, \quad
\chi=\sqrt{\ds{\frac{4s^2\tau+\sigma}{4s^2}}}.
\end{array}
\end{gather*}
Тогда на любом совместном уровне интегралов в пересечении с
$\mathcal{M}_3$ имеем
\begin{gather}\label{en4_4}
\begin{array}{l} \displaystyle{\alpha_1=\frac{(\mathcal{A}-r^2 U_1
U_2)(4 s^2 \tau+U_1 U_2)-(\tau+r^2) M_1 N_1 M_2 N_2 V_1 V_2}{4 r^2
s\, \tau (U_1+U_2)^2},
} \\[3mm]
\displaystyle{\alpha_2=\ri \frac{(\mathcal{A} -r^2 U_1 U_2)V_1 V_2
-(4 s^2 \tau+U_1 U_2)(\tau+r^2) M_1 N_1 M_2 N_2}{4 r^2 s\, \tau
(U_1+U_2)^2},
}\\[3mm]
\displaystyle{\beta_1=\ri \frac{(\mathcal{B}+r^2 U_1 U_2)V_1 V_2-(4
s^2 \tau+U_1 U_2)(\tau-r^2) M_1 N_1 M_2 N_2}{4 r^2 s\, \tau
(U_1+U_2)^2},
} \\[3mm]
\displaystyle{\beta_2=-\frac{(\mathcal{B} + r^2 U_1 U_2)(4 s^2 \tau
+ U_1 U_2)-(\tau-r^2) M_1 N_1 M_2 N_2 V_1 V_2}{4 r^2 s\, \tau
(U_1+U_2)^2},
}\\[3mm]
\displaystyle{\alpha_3= \frac{R }{r \sqrt{\mathstrut 2} } \, \frac
{M_1 M_2}{u_1+u_2},}\quad \displaystyle{\beta_3= - \ri \frac{R}{r
\sqrt{\mathstrut 2} } \, \frac { N_1 N_2}{u_1+u_2},}
\\[3mm]
\displaystyle{\omega_1=  \frac{R }{4 r s\, \sqrt{s \,\tau}} \,
\frac{M_2 N_1
U_1 V_2 + M_1 N_2 U_2 V_1}{ u_1^2-u_2^2},} \\[3mm]
\displaystyle{\omega_2= -\frac{\ri\,R}{4 r s\, \sqrt{s \,\tau} } \,
\frac{M_2 N_1 U_2 V_1 + M_1 N_2 U_1 V_2}{u_1^2-u_2^2},} \\[3mm]
\displaystyle{\omega_3= \frac{U_1-U_2}{\sqrt{2s\tau}}\frac{M_2 N_2
V_1 - M_1 N_1 V_2}{u_1^2-u_2^2}.}
\end{array}
\end{gather}
Здесь обозначено
\begin{gather*}
\begin{array}{l}
U_1=\sqrt{\mstrut u_1^2-\sigma}\,, \quad U_2=\sqrt{\mstrut
u_2^2-\sigma}\,, \quad V_1 = \sqrt{\mstrut 4s^2\chi^2-u_1^2}\,,
\quad V_2 = \sqrt{\mstrut
4s^2\chi^2-u_2^2}\,,\\[3mm]
M_1=\sqrt{\mstrut u_1+\tau+r^2}\,, \quad M_2=\sqrt{\mstrut
u_2+\tau+r^2}\,, \\[3mm]
N_1=\sqrt{\mstrut u_1+\tau-r^2}\,, \quad
N_2=\sqrt{\mstrut u_2+\tau-r^2}\,,\\[3mm]
R=\sqrt{\mstrut u_1 u_2 +\sigma^2+ U_1 U_2}\,,\\[3mm]
\mathcal{A}=[(u_1+\tau+r^2)(u_2+\tau+r^2)-2(p^2+r^2)r^2]\tau,\\[3mm]
\mathcal{B}=[(u_1+\tau-r^2)(u_2+\tau-r^2)+2(p^2-r^2)r^2]\tau.
\end{array}
\end{gather*}

Заметим, что пересечение множеств $\Delta_1$ и $\pov_3$ возможно
только в точках кривых $\delta_1 - \delta_3,$ которым отвечают
критические точки ранга 1. Образ $\Delta_2$ на $\pov_3$ задается
уравнением ${\tau=0}.$ Предельная форма соответствующих выражений
для фазовых переменных указана в \cite{Kh2009}. Геометрически
$\Delta_1$ и $\Delta_2$ являются линиями касания, соответственно,
$\pov_1$ с $\pov_2$ и $\pov_2$ с $\pov_3.$ Следует ожидать, что
ребро возврата, имеющееся на поверхности $\pov_3$ и заданное
параметрическими уравнениями
\begin{gather*}
\Delta_3 =\left\{h=\ds{\frac{3s^4+a^2b^2}{2 s^3}},\,
k=\ds{-\frac{3s^2}{4}+ p^2 -\frac{3a^2 b^2}{2 s^2}+\frac{a^4 b^4}{4
s^6}}, \, g=\ds{\frac{s^4+3a^2 b^2}{2s}} \right\},
\end{gather*}
также будет отвечать вырожденным точкам.
\begin{teo}{\it За исключением точек, лежащих в прообразе кривых $\Delta_2,
\Delta_3,$ все критические точки ранга 2 на многообразии
$\mathcal{M}_3$ невырождены. Они имеют тип $(2,1,0,0),$ если
значение $\tau s [s^4-(a^2+b^2-\tau)s^2+a^2 b^2 ]$ отрицательно, и
$(2,0,1,0),$ если оно положительно.}
\end{teo}
\doc.
В данном случае в качестве интеграла, имеющего особенность на
$\mathcal{M}_3,$ аналогично тому, как это сделано на
$\mathcal{M}_2,$ можно взять функцию, полученную исключением $s,
\tau$ из уравнений поверхности $\pov_3.$ Однако результат получается
слишком громоздким, и такой подход нерационален. Здесь удобно
рассмотреть функцию с неопределенными множителями Лагранжа, которая
введена в \cite{Kh34} для вывода уравнений критических подсистем,
\begin{gather*}
\Psi = 2 G- (p^2-\tau) H+s K.
\end{gather*}
При вычислении характеристического многочлена оператора $A_\Psi$
считаем $s,\tau$ константами, а затем подставляем в найденное
выражение значения (\ref{en4_4}). Получим
\begin{gather*}
\mu^2-\frac{2\tau}{s}\Bigl[s^4-(p^2-\tau)s^2+\frac{p^4-r^4}{4}\Bigr]=0.
\end{gather*}
Кратный корень (нулевой) имеется либо при $\tau=0$, что
соответствует множеству $\Delta_2$, либо при
$$
\tau=p^2-\ds{\frac{s^4+a^2b^2}{s^2}},
$$
что при подстановке в уравнения (\ref{en4_1}) для $\pov_3$ приводит
к кривой $\Delta_3.$  При отсутствии кратного корня тип точки
определяется знаком величины $\mu^2.$ \hfill $\square$

Таким образом, выполнена полная классификация особых точек
отображения момента неприводимой интегрируемой гамильтоновой системы
с тремя степенями свободы --- задачи о движении волчка типа
Ковалевской в двойном силовом поле, интегрируемость которой
установлена А.Г.\,Рейманом и М.А.\,Семеновым-Тян-Шанским.
Предъявлены явные формулы характеристических уравнений для
собственных чисел соответствующих симплектических операторов,
которые и определяют тип невырожденной особенности. Следующим этапом
является полное описание трехмерной топологии слоения Лиувилля
рассматриваемой системы.

\vspace{3ex}

\small

\makeatletter \@addtoreset{gather}{section}
\@addtoreset{footnote}{section}
\renewcommand{\section}{\@startsection{section}{1}{0pt}{1.3ex
plus 1ex minus 1ex}{1.3ex plus .1ex}{}}

{ 

\renewcommand{\refname}{{\rm\centerline{СПИСОК ЛИТЕРАТУРЫ}}}


\titleeng

\end{document}